\documentstyle[emulateapj,epsf]{article}

\lefthead{Lubow and Ogilvie}
\righthead{3D Waves in Disks}

\newcommand {\apgt} {\ {\raise-.5ex\hbox{$\buildrel>\over\sim$}}\ }
\newcommand {\aplt} {\ {\raise-.5ex\hbox{$\buildrel<\over\sim$}}\ }

\begin{document}

\title{Three-Dimensional Waves Generated at Lindblad Resonances in
Thermally Stratified Disks}

\author{S. H. Lubow and G. I. Ogilvie\altaffilmark{1}\\ Space
Telescope Science Institute\\ 3700 San Martin Drive\\ Baltimore, MD
21218
%\\ lubow@stsci.edu gogilvie@ast.cam.ac.uk
}
\altaffiltext{1}{Present 
address: Institute of Astronomy,
University of Cambridge, Madingley Road, Cambridge CB3~0HA, United
Kingdom.}

\slugcomment{Accepted for publication in the Astrophysical Journal}

\begin{abstract} 

We analyze the linear, 3D response to tidal forcing of a disk that is
thin and thermally stratified in the direction normal to the disk
plane.  We model the vertical disk structure locally as a polytrope
which represents a disk of high optical depth.  We solve the 3D
gas-dynamic equations semi-analytically in the neighborhood of a
Lindblad resonance.  These solutions match asymptotically on to those
valid away from resonances (previously obtained by Korycansky \&
Pringle 1995) and provide solutions valid at all radii $r$.  We obtain
the following results.  1) A variety of waves are launched at
resonance, including r modes and g modes. However, the f mode carries
more than 95\% of the torque exerted at the resonance.  2) These 3D
waves collectively transport exactly the amount of angular momentum
predicted by the standard 2D resonant torque formula.  3) Near
resonance, the f mode behaves compressibly and occupies the full
vertical extent of the disk. Away from resonance, the f mode behaves
incompressibly, becomes confined near the surface of the disk, and, in
the absence of other dissipation mechanisms, damps via shocks.  In
general, the radial length scale for this process is roughly $r_{\rm
L}/m$ (for resonant radius $r_{\rm L}$ and azimuthal tidal forcing
wavenumber $m$), {\it independent of the disk thickness $H$}. This
wave channeling process is due to the variations of physical
quantities in $r$ and is not due to wave refraction.  4) However, the
inwardly propagating f mode launched from an $m=2$ inner Lindblad
resonance experiences relatively minor channeling
(accompanied by about a factor of 5 increase in nonlinearity), all the way to the radial
center of the disk.

We conclude that for binary stars, tidally generated waves at Lindblad
resonances in highly optically thick circumbinary disks are subject to
strong nonlinear damping by the channeling mechanism, while those in
circumstellar accretion disks are subject to weaker nonlinear
effects.  We also apply our results to
waves excited by young planets for which $m \approx r/H$ and conclude
that the waves are damped on the scale of a few $H$.

\end{abstract}
 

\section{Introduction}

Gaseous disks are found in many types of binary star systems,
including cataclysmic variables (CVs) and pre-main-sequence stars, and
young planetary systems.  The orbiting objects (stars or planets)
exert tidal forces on these disks, which generally act merely to
distort the disks from an axisymmetric form. However, at special
locations in disks where resonances occur, the tidal forces generate
waves that transport energy and angular momentum.  As a result, a
resonant torque is exerted by the system objects.  The orbital
evolution of the perturbing objects sometimes depends on the strength
of these torques (Goldreich \& Tremaine 1980, hereafter GT80; Lin \&
Papaloizou 1993; Lubow \& Artymowicz 1996).  The waves transfer their
angular momentum and energy to the disk in the regions of space where
they damp, and this in turn affects the evolution of the disk.  For
example, gaps could be created in disks in regions where the waves
damp.

This paper concentrates on Lindblad resonances (LRs), which are due to
horizontal forcing (along the disk plane).  We assume throughout that
the disk is coplanar with the orbit of the system objects.  A 2D,
linear theory for resonant tidal torques and associated wave
propagation was developed by Goldreich \& Tremaine (1979, hereafter
GT79). This theory provides an explicit formula for the Lindblad
resonant torque.  The 2D theory considers the disk to have only radial
and azimuthal extent and ignores effects over its vertical extent
(perpendicular to the orbit plane).  Important progress has been made
in the study of 2D nonlinear waves in disks (Shu, Yuan, \& Lissauer
1985; Yuan \& Cassen 1994). The torque in the nonlinear case was found
to be within a few percent of that predicted by the 2D linear
formula. The nonlinearities produce highly spiked density profiles
which increase the level of dissipation present in a viscous disk.
Radiative damping of linear waves can be important particularly when
the disk is warm (Cassen \& Woolum 1996). Turbulent viscosity
in the disk provides another means of wave damping.

The 2D treatment is valid if both the vertical structure of the disk
and its thermodynamic response are locally isothermal.  Under such
circumstances, tidal forcing will generate a 2D wave in a 3D disk. The
wave front remains perpendicular to the disk plane at all heights, as
the wave propagates radially in the disk.  However, this 2D wave is
highly singular in that it does not exist in a disk with a vertical
temperature variation (Lin, Papaloizou, \& Savonije 1990a, hereafter
LPS; Lubow \& Pringle 1993, hereafter LP).  Furthermore, a vertically
isothermal structure is not realistic for many important classes of
gaseous disks, such as accretion disks in CVs, circumstellar and
circumbinary disks of YSOs, and protoplanetary disks. Such disks often
have optical depths much greater than unity and can be expected to
have substantial vertical temperature variations, if they have an
internal heat source such as turbulent dissipation.

We demonstrate in this paper that 2D tidal forcing, caused by LRs,
excites 3D waves in a thermally stratified disk.  However, another
class of resonances exist due to 3D effects. These resonances are due
to vertical, tidal forcing by a coplanar perturber of a disk with
nonzero thickness (Lubow 1981).  The vertical resonances also generate
horizontally propagating waves. Although intrinsically weaker than the
LRs, the vertical resonances may be of importance in close binary star
systems.

Some investigations of 3D effects have been carried out using
numerical simulations (Lin, Papaloizou, \& Savonije 1990a,b).  Such
approaches can be used to explore a limited range of physical
parameter space.
% and may be subject to unwanted effects of viscosity,
% which is used to stabilize the method.
However, recent progress has been made in obtaining semi-analytic
solutions for waves in 3D disks (LP; Korycansky \& Pringle 1995,
hereafter KP). The aim of this paper is to extend that approach to
understand 3D wave generation at LRs and the subsequent wave
propagation.  We determine the linear response of a thin (but nonzero
thickness) disk.  We model the disk locally as a polytrope in the
vertical direction, which is valid for a disk of very high optical
depth, and we ignore the effects of atmospheric layers.  We aim to
understand which modes of a thermally stratified disk are excited at
LRs and how much torque is carried by such waves. By studying the
properties of linear wave propagation, we can also understand where
nonlinearity sets in that will likely lead to shocks and subsequent
wave dissipation.

The outline of this paper is as follows. In \S2, we review the
properties of disk modes. In \S3 and \S4, we derive and solve the
equations for waves generated at LRs.  In \S5, we compute the total
torque carried by these waves and determine which modes are excited.
In \S6, we summarize properties of the f mode of the disk and present
our numerical results.  In \S7, we discuss the application of our
results to binary and protoplanetary systems. \S8 contains a summary.

\section{Review of three-dimensional waves in a thin disk}

\subsection{Free Waves}

In a thin disk, there is a separation of scales between the horizontal
and vertical directions which implies that the waves take a WKB form
in the radial direction, except in the neighborhood of resonances,
which are also turning points for the waves.  
The equations governing
axisymmetric WKB waves in a vertically isothermal disk are described
in LP within the approximation of the shearing sheet, in which the
radial derivatives of all base-state quantities other than the angular
velocity are neglected.  The case of a vertically polytropic disk was
subsequently considered by KP.

The problem reduces to a second-order system of ordinary differential
equations in the vertical coordinate $z$, which, together with
appropriate boundary conditions, constitutes an eigenvalue problem for
the frequency $\omega$ of the mode.  The radial wavenumber $k(r)$
appears as a parameter, so that a local dispersion relation
$\omega(k;r)$ is defined.  Moreover, the same dispersion relation
applies to non-axisymmetric waves if the frequency eigenvalue $\omega$
is replaced by the intrinsic frequency $\hat\omega=\omega-m\Omega$,
provided that the azimuthal wavenumber $m > 0$ satisfies $m \ll r/H$.

The solutions of the eigenvalue problem can be classified by analogy
with the theory of stellar oscillations (e.g.~Cox 1980).  The name of
each class of mode refers to the dominant restoring force that is
involved in its dynamics.  Thus p modes are acoustic modes, which are
inherently compressible and propagate by pressure forces.  The g modes
are gravity modes, which rely on buoyancy forces resulting from an
entropy gradient in the vertical direction.  There are also r modes,
which do not exist in a non-rotating star, and which propagate by
inertial forces.  The p, g, and r modes can be further classified
according to their symmetry about the mid-plane and the number
of nodes in the pressure perturbation (or radial velocity
perturbation) in $z>0$.  Finally, the f mode is the fundamental mode
of oscillation, and has unique characteristics described in detail in
\S6.1 below.  In fact, a disk has two f modes, one of each symmetry,
because it has two surfaces.

Some of these modes are lost in special cases.  In an incompressible
disk, there are no p modes, although all other modes survive.  The g
modes do not exist in a disk that is adiabatically stratified in the
vertical direction.  Similarly, there are no r modes in a disk that
has zero epicyclic frequency.  In an isothermal disk with no surface,
the g modes are lost and the f mode of even symmetry takes a special
form, known otherwise as the Lamb mode or two-dimensional mode. (There
has been some confusion in terminology. LP used the naming convention
that r modes were called g modes.)

For a monochromatic wave or a normal mode of the disk, $\omega$ is
independent of $r$ and the dispersion relation can be followed
continuously to determine the wavenumber at any radius.  Typically,
the wave occupies a finite interval in radius, which is bounded either
by turning points, at which $k$ goes to zero, or by the edges of the
disk.  Moreover, the continuous variation of the amplitude of the wave
with $r$ can be determined by appealing to the conservation of either
energy or angular momentum wave action.

In many cases of interest, the dispersion relation is followed into a
limit in which the dimensionless wavenumber $kH$ becomes large.  The
behavior of the modes in this limit was described by Ogilvie (1998).
Provided that the disk has a surface, the f, p, and g modes become
confined in a layer of characteristic thickness $k^{-1}$ near the
surface, and their frequencies scale proportionally to $k^{1/2}$.  In
contrast, the r modes become confined in a layer of characteristic
thickness $k^{-1/2}H^{1/2}$ centered on the mid-plane, and
their frequencies scale proportionally to $k^{-1/2}$, unless the disk
is marginally stable to convection.  None of the modes occupies the
full vertical extent of the disk in this limit.

It is useful to note the following properties of waves in a Keplerian
disk.  The f and g modes have $|\hat\omega|\ge\Omega$ and propagate
both inside the inner LR and outside the outer LR.  The r modes have
$|\hat\omega|\le\Omega$ and propagate in the region between the two
LRs, where the corotation resonance is also located.  An exception is
the `tilt' mode, consisting of the union of the f and ${\rm r}_1$
modes of odd symmetry, which propagates at all radii.  The p modes
have $|\hat\omega|\ge\omega_{\rm p}$, where $\omega_{\rm p}>\Omega$ is
a frequency which increases with the order of the mode.  Their
resonances are separated from the Lindblad resonances by a distance
which also increases with the order of the mode.  These properties are
summarized in Figure~1.

\placefigure{fig1}

\subsection{Asymptotic Matching}

One of the goals of this paper is to determine to level of excitation
of the free modes described above, due to tidal forcing at Lindblad
resonances. To accomplish this, we determine in \S4 the solutions for
the tidally generated waves in a small region close to the Lindblad
resonance (of order $(H^2 r)^{1/3}$ in radius), where the wave
excitation occurs.  In this region of space, a locally valid
approximation is used, in which the quantity $\kappa^2-\hat\omega^2$,
which has a zero at the resonance, is treated as a linear function of
radius.  Away from the resonance (over most of space), a different
approximation applies, which is a WKB approximation in the radial
direction.  Radial WKB free wave solutions of a vertically polytropic
disk were obtained by KP. A solution that is uniformly valid for all
$r$ (asymptotically accurate in all space in the limit $H \ll r$) is
obtained by the method of matched asymptotic expansions (e.g. van Dyke
1975).  In this context, the global wave solutions are determined by
matching the outer limit (limit going away from resonance) of the
driven Lindblad resonance solutions to the inner limit (limit
approaching resonance) of the KP free waves. In practice, the level of
excitation of each KP mode is easily determined from the amount of
angular momentum generated for that mode in the region near resonance.

\section{Mathematical formulation}

\subsection{Base state}

Consider a cylindrical coordinate system $(r, \phi, z)$ such that
$z=0$ coincides with the disk mid-plane and the direction of
increasing $\phi$ is the direction of disk rotation.  The unperturbed
disk has angular velocity $\Omega(r)$, density $\rho(r,z)$, and
pressure $p(r,z)$. We assume that the disk is thin and ignore effects
of viscosity and self-gravity.  As in KP, we adopt a polytropic disk
structure in $z$ locally in $r$ so that
\begin{equation}
\rho(r,z) = \rho_0(r) (1 - z^2/H^2)^s
\end{equation}
and
\begin{equation}
p(r,z) = p_0(r) (1 - z^2/H^2)^{s+1},
\end{equation}
where $\rho_0(r)$ and $p_0(r)$ are arbitrary smooth functions of $r$
and $s$ is the polytropic index. The disk half-thickness $H(r)$
satisfies
\begin{equation}
H^2(r) = \frac{2(s+1) p_0(r)}{\Omega^2(r) \rho_0(r)}.
\end{equation}
The thin-disk approximation used to derive the above requires $H \ll
r$.  The base-state angular velocity is slightly modified from a
Keplerian law by the radial pressure forces and it varies slightly in
$z$ within the disk.  However, these effects are unimportant for the
issues described in this paper and are generally ignored.  The
polytropic structure provides a mathematically convenient description
of a thermally stratified disk of high optical depth. However, the
features of the results obtained in subsequent sections do not depend
on the thermally stratified disk being polytropic.

\subsection{Dynamical equations}

Let the local Eulerian perturbations of disk velocity in cylindrical
coordinates be represented by $(u, v, w)$, that of pressure by
$p^{\prime}$, and that of density by $\rho^{\prime}$.  We represent
physical wave quantities such as the pressure perturbation by
$p^\prime(r, \phi, z, t) = p^{\prime}(r,z) \exp{[im (\phi -
\Omega_{\rm p} t)]}$, where positive integer $m$ is the azimuthal
wavenumber and $\Omega_{\rm p}$ is the angular pattern speed of the
wave.  Consider the disk to be subject to a tidal potential component
$\Phi_m(r,\phi,z,t)$ of the above form.  In the case that the system
objects are in circular orbits, the pattern speed is equal to the
angular orbital speed of the objects. If the system has eccentricity,
then additional components of the perturbing potential appear
containing the pattern speeds that differ from the fundamental orbital
frequency of the system (see GT80).  Using the 3D shearing sheet
equations (see LP and KP), we obtain the following equations for the
disk perturbations.
\begin{equation}
\label{u}
-i \hat{\omega} u - 2 \Omega v + \frac{\partial_{r} p^{\prime}}{\rho}
= -\frac{d \Phi_m}{dr},
\end{equation}
\begin{equation}
\label{v}
-i \hat{\omega} v + 2 B u = -\frac{im}{r} \Phi_m,
\end{equation}
\begin{equation}
\label{w}
-i \hat{\omega} w + g \frac{\rho^{\prime}}{\rho} +
\frac{\partial_{z} p^{\prime}}{\rho} = 0,
\end{equation}
\begin{equation}
\label{mc}
-i \hat{\omega} \frac{\rho^{\prime}}{\rho} + w \partial_{z}
\ln\rho + \partial_{r} u + \partial_{z} w =0,
\end{equation}
and
\begin{equation}
\label{en}
-i \hat{\omega} \left(\frac{p^{\prime}}{p} -
\gamma \frac{\rho^{\prime}}{\rho}\right) + w \partial_z
\ln{\left(\frac{p}{\rho^{\gamma}}\right)} = 0,
\end{equation}
where $\gamma$ is the usual adiabatic exponent.  These equations are
respectively the radial, azimuthal, and vertical force equations, the
mass conservation equation, and the energy equation for adiabatic
perturbations.  The frequency $\hat{\omega}$ is defined as
\begin{equation}
\hat{\omega}(r) = m\left[\Omega_{\rm p} - \Omega(r)\right].
\end{equation}
Quantity $B(r)$ is the usual Oort constant equal to $\Omega +
\frac{r}{2} \frac{d \Omega}{ dr}$ and $g$ is the vertical gravity in
the disk equal to $g(r,z) = \Omega^2(r) z$.  The azimuthal pressure
force $i m p^{\prime}/r$ was ignored in equation (\ref{v}) and the
azimuthal velocity term $i m v/r$ was ignored in equation (\ref{mc}).
This approximation is justified for a thin-disk limit for which
\begin{equation}
\label{td}
 H \ll r/m.
\end{equation}
The forcing term in the vertical force equation (\ref{w}) was
ignored. This approximation can be justified by the thin-disk
condition (\ref{td}) everywhere except at vertical resonance points
(Lubow 1981). We concentrate on wave properties near Lindblad
resonances (horizontal resonances) in this paper and thus this
approximation is justified.

\section{Equations near LRs}

By combining equations (\ref{u}) and (\ref{v}), we obtain
\begin{equation}
\label{u1}
u(\hat{\omega}^2-\kappa^2) + i
\hat{\omega} \frac{\partial_r p^{\prime}}{\rho} = 
-i \hat{\omega} \frac{d \Phi_m}{dr} + \frac{ 2 i m \Omega}{r}
\Phi_m,
\end{equation}
where $\kappa$ is the epicyclic frequency such that $\kappa^2 = 4B
\Omega$.  Notice that equations (\ref{w})--(\ref{en}), together with
equation (\ref{u1}), form a complete set, i.e., $v$ has been
eliminated. Once $u$ has been determined, $v$ can be obtained from
equation (\ref{v}).

LRs occur at radii $r_{\rm L}$ where $\hat{\omega}^2 = \kappa^2$.
Away from such locations, the above equations are equivalent to the
WKB equations of KP.  Near an LR, the WKB approximation breaks down in
equation (\ref{u1}), but the solutions can be represented in terms of
Airy functions in the neighborhood of the turning point.  To treat
this region, we follow standard techniques (e.g. GT79) and expand in
the neighborhood of the resonance
\begin{equation}
\label{D}
  \kappa^2 - \hat{\omega}^2 \approx {\cal D} x,
\end{equation}
where $x= (r - r_{\rm L})/r_{\rm L}$, and ${\cal D} = r d (\kappa^2 -
\hat{\omega}^2)/dr$ is evaluated at $r=r_{\rm L}$.  We then
approximate the radial force equation by
\begin{equation}
\label{u2}
- u {\cal D} x + i
\hat{\omega} \frac{\partial_r p^{\prime}}{\rho} = 
-i \hat{\omega} \frac{d \Phi_m}{dr} + \frac{ 2 i m \Omega}{r}
\Phi_m.
\end{equation}

\subsection{Free waves}

The equations for free waves near LRs are obtained by setting
$\Phi_m=0$ and combining equations (\ref{w})--(\ref{en}) and (\ref
{u2}).  We seek solutions of the following form.
\begin{equation}
\label{ff}
\begin{array}{lcl}
u(x, z) & = & Ai(q x) \tilde{u}(z),\\ v(x, z) & = & i Ai(q x)
\tilde{v}(z),\\ w(x, z) & = & q Ai^{\prime}(q x) \tilde{w}(z),\\
p^{\prime}(x,z) & = & iq Ai^{\prime}(q x) \tilde{p}^{\prime}(z),\\
\rho^{\prime}(x,z) & = & iq Ai^{\prime}(q x) \tilde{\rho}^{\prime}(z),\\
\end{array}
\end{equation}
where $Ai$ is the Airy function given by equation (10.4.1) of
Abramowitz \& Stegun (1965) and $Ai^{\prime}$ is the derivative with
respect to its argument. The parameter $q$ is a dimensionless constant
that will be determined as an eigenvalue.  (Solutions involving the
complementary Airy function $Bi$ have been discarded because, on one
side of resonance, they approach infinity with distance from the
resonance.)

Using the property that $ Ai^{\prime \prime}(x) = x Ai(x)$, we obtain
the following free-wave equations for $\tilde{u}, \tilde{v},
\tilde{w}, \tilde{p}^{\prime}$ and $\tilde{\rho}^{\prime}$.
\begin{equation}
\label{ufw}
\rho {\cal D} \tilde{u} + \frac{\hat\omega q^3}{r} \tilde{p}^{\prime} = 0,
\end{equation}
\begin{equation}
\label{vfw}
\tilde{v} = -\frac{2 B}{\hat{\omega}} \tilde{u},
\end{equation}
\begin{equation}
\label{wfw}
\hat{\omega} \rho \tilde{w} - g \tilde{\rho}^{\prime} -
\frac{d \tilde{p}^{\prime}}{dz} = 0,
\end{equation}
\begin{equation}
\label{mcfw}
\hat{\omega} \tilde{\rho}^{\prime} +
\tilde{w} \frac{d\rho}{dz} + \rho \left(\frac{\tilde{u}}{r}
 + \frac{d \tilde{w}}{dz}\right) =0,
\end{equation}
and
\begin{equation}
\label{enfw}
\hat{\omega} \left(\frac{\tilde{p}^{\prime}}{p} -
\gamma \frac{\tilde{\rho}^{\prime}}{\rho} \right) +  w \frac{d}{dz}
\ln{\left(\frac{p}{\rho^{\gamma}}\right)} = 0,
\end{equation}
where all coefficients are evaluated at $r=r_{\rm L}$.

As can be seen from the above, the equations have been reduced to 1D
equations in $z$.  We introduce dimensionless variables
\begin{equation}
\label{nd}
\begin{array}{lcl}
Z & = & z / H,\\ X(Z) & = & \hat{\omega} \tilde p^{\prime}(z)/(
\Omega^3 H^2 \rho), \\ F & = & \hat{\omega}/\Omega, \\ Q & = &
(H/r)^{2/3} q,\\ S & = & {\cal D} / \Omega^2,\\ W(Z) & = & \tilde
w(z)/(\Omega H).\\
\end{array}
\end{equation}
Combining the above equations (\ref{ufw})--(\ref{nd}), we obtain
\begin{equation}
\label{Xfw}
\frac{d X}{d Z} - \left( \frac{ 2 \alpha Z}
{1-Z^2} \right) X - \left(F^2 - \frac{2 \alpha Z^2} {1-Z^2} \right) W =
0
\end{equation}
and
\begin{equation}
\label{Wfw}
\frac{d W}{d Z} - \left( \frac{ 2 \beta Z}
{1-Z^2} \right) W + \left(\frac{2 \beta}{1-Z^2} - \frac{Q^3}{S} \right)
X = 0,
\end{equation}
where the constants $\alpha = s - (1+s)/\gamma$ and
$\beta=(1+s)/\gamma$ are proportional to the square of the disk's
buoyancy frequency and the square of the inverse sound speed at
mid-plane, respectively.  We note that these equations can be derived
from equations (13) and (14) of KP in an expansion about the point of
the dispersion relation where $k=0$ and $F^2=\kappa^2$.

Equations (\ref{Xfw}) and (\ref{Wfw}) are integrated numerically with
appropriate boundary conditions at $Z=0$ and $Z=1$, as in KP.  Both
even and odd solutions about $Z=0$ can be obtained, but for the
coplanar forcing problem at hand, only even solutions are relevant.
Such solutions must then satisfy $W(0) = 0$.  Infinitely many
solutions exist which differ in $n$, the number of vertical nodes of
$X$ (the pressure perturbation) for $Z>0$.  At $Z=1$, the boundary
condition is that the Lagrangian pressure perturbation must
vanish. Solutions are obtained by numerically integrating out of the
regular singular point $Z=1$ down to $Z=0$.  Only certain values of
$Q$ permit the mid-plane boundary condition $W=0$ to be satisfied.
Once the value of $Q$ is determined, the spatial structure of all
physical quantities is known. Solutions with various spatial
structures are obtained that behave as g modes, r modes, and the f
mode.  Each solution is characterized the type of mode (g, r, etc.)
and by the value of $n$.  There are a countably infinite number of
solutions, which we label by index $j$.

We note that the scalings for a thin disk are such that $q$ is of
order $(r/H)^{2/3}$.  The characteristic width of the resonance is
therefore $(H^2r)^{1/3}$, intermediate between $H$ and $r$.  In
addition, the vertical component of the velocity is smaller than the
horizontal components by a factor of order $(H/r)^{1/3}$.

\subsection{Driven waves}

We return to the problem of wave generation due to tidal forcing.  We
consider solutions to the inhomogeneous equations near resonance,
equations (\ref{w})--(\ref{en}) and (\ref{u2}).  We consider the set
of all homogeneous solutions obtained in the last section, each of
which is characterized by the set of physical quantities $q_j$,
$\tilde u_j(z)$, $\tilde v_j(z)$, $\tilde w_j(z)$, $\tilde
p^\prime_j(z)$, and $\tilde\rho^\prime_j(z)$.  We construct possible
solutions to the inhomogeneous equations which are expanded in terms
of the homogeneous solutions in the following way.
\begin{equation}
\label{frf}
\begin{array}{lcl}
u(x, z) & = & \sum_j a_j\tilde u_j(z) \left[ Ai(q_j x) + i s_j Gi(q_j
x) \right],\\ w(x, z) & = & \sum_j a_j q_j\tilde w_j(z) \left[
Ai^{\prime}(q_j x) + i s_j Gi^{\prime} (q_j x) \right],\\
p^{\prime}(x,z) & = & i \sum_j a_j q_j\tilde p^\prime_j(z) \left[
Ai^{\prime}(q_j x) + i s_j Gi^{\prime} (q_j x) \right],\\
\rho^{\prime}(x,z) & = & i \sum_j a_j q_j\tilde\rho^\prime_j(z) \left[
Ai^{\prime}(q_j x) + i s_j Gi^{\prime} (q_j x) \right],\\
\end{array}
\end{equation}
where $Gi$ is the inhomogeneous Airy function defined in equation
(10.4.55) of Abramowitz \& Stegun (1965), $s_j$ is either $+1$ or
$-1$, and $a_j$ are the expansion coefficients.  The linear
combination of Airy functions, $Ai(q_j x) + i s_j Gi(q_j x) $ and
$Ai^{\prime}(q_j x) + i s_j Gi^{\prime} $, in the above is chosen to
provide traveling waves in the limit $q_j x \rightarrow - \infty$.
The sign of $s_j$ in the above determines the radial propagation
direction of the wave (toward or away from the resonance) which we
leave as an open issue for now for each $j$.

We use the fact that $Gi$ satisfies $ Gi^{\prime \prime}(x) - x Gi(x)
= -1/\pi$, and substitute the above forms into inhomogeneous equations
(\ref{w})--(\ref{en}) and (\ref{u2}).  We find that equations
(\ref{w})--(\ref{en}) are automatically satisfied and do not in any
way constrain the expansion coefficients. The constraint comes from
the force equation (\ref{u2}), which is satisfied provided that
\begin{equation} 
\label{a_j}
\Psi_m
 = -\frac{1}{\pi} \sum_j s_j a_j q_j^2
 \frac{\tilde p^\prime_j(z)}{\rho(z)},
\end{equation}
where forcing term $\Psi_m$ is defined by
\begin{equation}
\Psi_m = r\frac{d \Phi_m}{dr} - \frac{ 2 m \Omega }{\hat{\omega}}
\Phi_m,
\end{equation}
evaluated at $r=r_{\rm L}$.  In equation (\ref{a_j}), the left-hand
side is independent of $z$.  In the case of an isothermal disk with
adiabatic index equal to unity ($\gamma=1$), the ratio $\tilde
p^\prime_j/\rho$ is independent of $z$ (and proportional to the sound
speed squared) for the 2D mode, which is then the proper solution in
that case only. In general, for disks with non-isothermal vertical
structures or vertically isothermal disks having $\gamma \ne 1$, the
quantity $\tilde p^\prime_j/\rho$ varies with $z$ for all modes in the
disk. So some suitable combination of solutions is required.

To solve equation (\ref{a_j}), we seek an appropriate inner product
that allows us to invert that equation to solve for $a_j$.  This can
be derived by noting that equations (\ref{ufw})--(\ref{enfw}) can be
combined to give a single equation for $\tilde u$ in the form
\begin{equation}
{{d}\over{dz}}\left(f_1{{d\tilde u}\over{dz}}\right)+f_2\tilde
u+\frac{\lambda\rho\tilde u}{H^2 \Omega^2 }=0,\label{qsl}
\end{equation}
where
\begin{equation}
f_1={{\rho}\over{N^2-\hat\omega^2}},
\end{equation}
\begin{equation}
\label{ldef}
\lambda={{q^3 H^2 \Omega^2 }\over{r^2{\cal D}}}
\end{equation}
is a dimensionless eigenvalue, 
and $N$ is the vertical buoyancy frequency in the disk.  (The detailed
form of $f_2$ is not illuminating.)  Equation (\ref{qsl}) does not
satisfy the conditions of the Sturm--Liouville theorem, because $f_1$
is not of definite sign and both $f_1$ and $f_2$ vanish at $z=\pm H$.
(We note in passing that the singular point where $\hat\omega^2=N^2$
is only an {\it apparent} singularity, since both linearly independent
solutions are in fact regular there.)  However, it can still be
demonstrated that the eigenvalues $\{\lambda_j\}$ are real and the
eigenfunctions $\{\tilde u_j\}$ orthogonal in the sense that
\begin{equation}
\int\rho\tilde u_j^*\tilde u_k\,dz=0
\end{equation}
for any two modes with distinct eigenvalues.  (Here and below,
integrals are taken over the full vertical extent of the disk.)  This
is a limiting case of a more general orthogonality relation,
\begin{equation}
\int\rho\left(\tilde u_j^*\tilde u_k+\tilde w_j^*\tilde w_k\right)\,dz=0,
\end{equation}
which applies to the eigenfunctions away from resonances studied by
KP.  Moreover, it can be asserted that the eigenfunctions form a
complete set on the space of continuous functions on $(-H,H)$,
although, technically, the pointwise convergence of an eigenfunction
expansion is non-uniform with respect to $z$ at the end-points where
the weight function $\rho$ vanishes.

By rewriting equation (\ref{a_j}) in the form
\begin{equation}
\Psi_m={{r{\cal
D}}\over{\pi\hat\omega}}\sum_j{{s_ja_j}\over{q_j}}\tilde u_j(z),
\end{equation}
and using the orthogonality relation, we obtain the coefficients as
\begin{equation}
\label{aj}
a_j={{\pi\hat\omega s_jq_j\Psi_m}\over{r{\cal D}}}\int\rho\tilde
u_j^*\,dz\bigg/\int\rho|\tilde u_j|^2\,dz.
\end{equation}

\section{Derivation of the torque formula}

We now proceed to show that the torque exerted by a companion object
at a Lindblad resonance is precisely equal to the value derived under
a two-dimensional approximation.  The radial velocity perturbation
corresponding to the collection of waves excited at the resonance is
known from \S4.2 to have an inner expansion
\begin{equation}
\label{uexp}
u(x,z) = \sum_ja_j\tilde
u_j(z)\left[Ai(q_jx)+is_jGi(q_jx)\right]\label{eq1}
\end{equation}
in the vicinity of the resonance, where $s_j=\pm1$ is still to be
determined, and a factor $\exp(-i\omega t+im\phi)$ is
understood.  From the asymptotic forms (Abramowitz \& Stegun 1965)
\begin{eqnarray}
x\to+\infty:&&Ai(x)\pm iGi(x)\sim\pm i\pi^{-1}x^{-1},\\
x\to-\infty:&&Ai(x)\pm iGi(x)\sim\pm
i\pi^{-1/2}(-x)^{-1/4}\nonumber\\
&&\qquad\times\exp\left\{\mp i\left[{\textstyle{{2}\over{3}}}
(-x)^{3/2}+{\textstyle{{1}\over{4}}}\pi\right]\right\},\label{aia}
\end{eqnarray}
it is clear that the component of the solution labeled $j$ matches on
to a WKB traveling wave in the region $q_jx<0$ and on to a decaying
disturbance in the region $q_jx>0$.

For single WKB waves of the disk it can be shown that the flux of
angular momentum wave action, averaged over $t$, integrated over
$\phi$ and $z$, and measured in the direction of increasing $r$, is
\begin{equation}
\label{Fa}
F^{\rm(a)}={{\pi
rm}\over{k}}\left({{\hat\omega^2-\kappa^2}
\over{\hat\omega^2}}\right)\int\rho|u|
^2\,dz.
\end{equation}
The density of angular momentum wave action, similarly averaged and
integrated, is
\begin{equation}
A^{\rm(a)}={{\pi
rm}\over{\hat\omega}}\int\rho\left(|u|^2+|w|^2\right)\,dz,
\end{equation}
and a straightforward application of first-order perturbation theory
confirms that the ratio $F^{\rm(a)}/A^{\rm(a)}$ is equal to the radial
group velocity
\begin{equation}
v_{\rm g}={{\partial\hat\omega}\over{\partial
k}}={{(\hat\omega^2-\kappa^2)}\over{\hat\omega
k}}\int\rho|u|^2\,dz\bigg/\int\rho\left(|u|^2+|w|^2\right)\,dz
\end{equation}
which is the slope of the WKB dispersion relation.  We therefore
identify the direction of propagation of a WKB wave according to the
sign of its radial group velocity.

Consider first the outer limit $x\to-\infty$ of equation (\ref{eq1})
for a component with $q_j>0$.  The wavenumber, computed as the
derivative of the phase, is
\begin{equation}
k\sim s_jq_j(-q_jx)^{1/2}r^{-1}.
\end{equation}
It follows that
\begin{equation}
{\rm sgn}(v_{\rm g})=s_j\,{\rm sgn}({\cal D}){\rm sgn}(\hat\omega),
\end{equation}
and, since ${\rm sgn}({\cal D})=-{\rm sgn}(\hat\omega)$ for a
Keplerian disk (with the convention that $m>0$), the outgoing wave
that is required by causality is the one with $s_j=+1$.  From
equations (\ref{uexp}), (\ref{aia}), and (\ref{Fa}), the angular
momentum flux associated with the wave is
\begin{equation}
F^{\rm(a)}_j={{r^2m{\cal D}|a_j|^2}\over{\hat\omega^2q_j^2}}\int\rho|\tilde
u_j|^2\,dz,
\end{equation}
although we emphasize that this is measured in the direction of
increasing $r$, while the wave propagates into $r<r_{\rm L}$.  By a
similar argument it can be shown that, for a component with $q_j<0$,
the required wave has $s_j=-1$ and carries an angular momentum flux
\begin{equation}
F^{\rm(a)}_j=-{{r^2m{\cal D}|a_j|^2}
\over{\hat\omega^2q_j^2}}\int\rho|\tilde u_j|^2\,dz
\end{equation}
into $r>r_{\rm L}$.  From equation (\ref{aj}), the torque exerted on
the disk is therefore
\begin{eqnarray}
T&=&-{{r^2m{\cal
D}}\over{\hat\omega^2}}\sum_j{{|a_j|^2}\over{q_j^2}}\int\rho|\tilde
u_j|^2\,dz\nonumber\\
&=&-{{\pi^2m\Psi_m^2}\over{\cal D}}\sum_j\left|\int\rho\tilde
u_j\,dz\right|^2\bigg/\int\rho|\tilde u_j|^2\,dz\nonumber\\
&=&-{{\pi^2m\Sigma_{\rm L}\Psi_m^2}\over{\cal D}},\label{T}
\end{eqnarray}
where $\Sigma_{\rm L}=\int\rho\,dz$ is the surface density at the
resonance, and we have made use of the completeness of the
eigenfunctions $\{\tilde u_j\}$.  This formula is in precise agreement
with equation (A10) of GT79.  This result is not surprising
since the 2D torque formula 
holds under a variety of physical
conditions (Goldreich \& Tremaine 1982; Papaloizou \& Lin 1984).
For non-Keplerian rotation curves, one
cannot rule out the possibility that ${\rm sgn}({\cal D})={\rm
sgn}(\hat\omega)$, and so a more general formula is
\begin{equation}
T=-{\rm sgn}(\hat\omega){{\pi^2m\Sigma_{\rm L}\Psi_m^2}\over{|{\cal D}|}}.
\end{equation}

We now introduce a specific disk model which is used for all the
numerical calculations in this paper.  The disk is Keplerian and its
vertical structure is polytropic, with index $s=3$.  The adiabatic
exponent is taken to be $\gamma=5/3$.  For this model we have computed
the fraction
\begin{equation}
f_j=\left|\int\rho\tilde
u_j\,dz\right|^2\bigg/\Sigma_{\rm L}\int\rho|\tilde u_j|^2\,dz
\end{equation}
of the torque carried by each mode.  The results for the f mode and
the first three g and r modes are presented in Table~1, together with
the corresponding eigenvalues $\lambda_j$ (defined by
eq. [\ref{ldef}])).  It is
clear that almost all the angular momentum is transported by the f
mode, although the r and g modes are also weakly excited.  These first
seven modes together account for 99.93\% of the torque.  For obvious
reasons we focus on the propagation of the f mode for the remainder of
this paper.  However, it is worth noting that the r modes propagate
away from the resonance in the opposite direction to the f and g
modes. The results in Table 1 are independent of $m$.

\begin{deluxetable}{lrr}
\tablecaption{Modes excited at a Lindblad resonance.\label{tab1}}
\tablewidth{0pt} \tablehead{\colhead{Mode}&\colhead{$\lambda_j\
$}&\colhead{$f_j$}} \startdata ${\rm
f}$&$6.16$&$0.9678$\nl ${\rm g}_1$&$112.26$&$0.0010$\nl ${\rm
g}_2$&$236.47$&$0.0003$\nl ${\rm g}_3$&$401.73$&$0.0002$\nl
$\dots$&$\dots$&$\dots$\nl ${\rm r}_1$&$-18.98$&$0.0277$\nl ${\rm
r}_2$&$-98.32$&$0.0018$\nl ${\rm r}_3$&$-237.42$&$0.0005$\nl
$\dots$&$\dots$&$\dots$\nl \enddata
\end{deluxetable}

\section{Propagation of modes in a Keplerian disk}

\subsection{Basic properties of the f mode}

In this section, we summarize the properties of the single most
important mode in this problem, the ${\rm f}^{\rm e}$ mode or
fundamental mode of even symmetry.  The eigenfunction of this mode,
whether at the resonance or away from it, is uniquely characterized as
follows.  The pressure perturbation (or radial velocity perturbation)
has no nodes in $-H<z<H$.  The vertical velocity perturbation has only
a single node at $z=0$, as required by symmetry.  This form of the
eigenfunction allows a large overlap integral with the forcing
potential which is independent of $z$, and explains why this mode is
excited more than any other (see eq. [\ref{aj}]).

The dispersion relation for this mode in our model disk (which is
Keplerian and polytropic, with $s=3$ and $\gamma=5/3$) is shown in
Figure~2a.  Also shown is the asymptotic form
\begin{equation}
\hat\omega^2\sim\Omega^2kH\label{eq2}
\end{equation}
derived by Ogilvie (1998), which is valid in the limit $kH\to\infty$.
Since the surface gravity of the disk is $g_{\rm s}=\Omega^2H$, the
above equation is consistent with the well-known dispersion relation
for surface waves in a deep layer of water.  A more accurate
approximation, which takes into account the effects of rotation and
stratification, is
\begin{equation}
\hat\omega^2\sim\Omega^2kH+{\textstyle{{1}\over{2}}}\kappa^2
-{\textstyle{{1}\over{2}}}s\Omega^2,\label{eq3}
\end{equation}
and this is shown in Figure~2b to provide an excellent approximation
even for values of $kH$ as small as 3.

\placefigure{fig2}

The form of the eigenfunction in this limit also corresponds to that
expected of a surface gravity mode.  For definiteness, we describe a
wave traveling radially outwards in the disk, but inward traveling
waves and standing waves are very similar in form.  The wave is
trapped in a layer of characteristic thickness $k^{-1}$ near the
surface of the disk, and in this layer the asymptotic forms
\begin{eqnarray}
u&\sim&A\cos(kr-\omega t+m\phi)\exp(-k\delta),\label{eq4}\\
w&\sim&A\sin(kr-\omega t+m\phi)\exp(-k\delta)\label{eq5}
\end{eqnarray}
are valid, where $\delta=H-|z|$.  This velocity field is both
irrotational and solenoidal (i.e.~non-compressive).  The azimuthal
velocity is smaller by a factor $O\left((kH)^{-1/2}\right)$.  In this
limit, the even and odd f modes are indistinguishable since there is a
loss of contact between the two surfaces of the disk (KP).

For small values of $kH$, the mode departs from the behavior described
above.  An approximation valid in the limit $kH\to0$ is
\begin{equation}
\hat\omega^2\sim\kappa^2+k^2 H^2 \Omega^2 /\lambda,
\end{equation}
where $\lambda>0$ is the dimensionless f-mode eigenvalue defined by equation
(\ref{ldef}).  This form is reminiscent of the dispersion relation
$\hat\omega^2=\kappa^2+c_{\rm s}^2k^2$ for the two-dimensional mode in
an isothermal disk (LP).  While it is true that $\lambda$ depends on
$\gamma$, implying that the ${\rm f}^{\rm e}$ mode is partly
compressive in this limit, nevertheless $\lambda$ has a non-trivial
limit when $\gamma\to\infty$.  Therefore the mode behaves partly
compressibly near the resonance, but only when the disk is
compressible.  The mode does not disappear in the incompressible
limit.

We emphasize how this description differs from earlier, less accurate
theories of waves in disks.  If the disk is described as a
two-dimensional fluid, then all radially propagating modes are
inherently compressive by virtue of the constraints on the fluid, and
would normally be described as sound waves modified by rotation.  In a
three-dimensional disk, a two-dimensional mode of this form exists
only if the disk is vertically isothermal.  In general, however, the
principal mode excited at an LR in a three-dimensional disk is the
${\rm f}^{\rm e}$ mode, which is not inherently compressive.
Moreover, this mode cannot be described using a vertical WKB
approximation (cf.~Vishniac \& Diamond 1989; Goodman 1993) because it
has no pressure nodes in the vertical direction, being, if anything, evanescent
rather than wavelike.

\subsection{Propagation and damping of the f mode}

\subsubsection{Remarks on the nonlinear development of the f mode}

If the waves excited at a resonance were able to propagate radially
without dissipation, they would reflect from the radial boundaries and
establish a standing wave pattern.  The net transfer of angular
momentum and energy between the disk and the companion would then be
zero.  In order for the waves to have a significant dynamical effect
on the evolution of the system, they must deposit their angular
momentum (either positive or negative) in the disk as a result of
being damped.  This can happen within linear theory if there is
sufficient viscosity, but otherwise nonlinear dissipation is required.
As noted by KP, the concentration of modes near the surface of a
polytropic disk is likely to result in enhanced dissipation through
the formation of shocks.

It might be argued that, since the f mode is approximately
non-compressive, it would form shocks less easily and would therefore
be less susceptible to this type of dissipation.  However, little is
known about the nonlinear development of surface gravity modes in a
compressible fluid.  In the case of an incompressible fluid, it is
well known that when the amplitude of a surface gravity wave is
increased, the distorted profile of the surface changes from a
sinusoidal shape to one with sharper crests, forming an angle
approaching $120^\circ$.  This happens when the velocity perturbation
is comparable to the phase velocity $(g_{\rm s}/k)^{1/2}$ (e.g.~Stokes
1880; Lamb 1932), and causes the crests to break.  Although we
consider it more probable that the nonlinear development in a
compressible fluid will lead to shocks rather than breaking crests, it
is still appropriate to normalize the solution described by equations
(\ref{eq4}) and (\ref{eq5}) according to
\begin{equation}
u\sim{\cal N}\left({{\Omega^2H}\over{k}}\right)^{1/2}
\cos(kr-\omega t+m\phi)\exp(-k\delta),
\end{equation}
where ${\cal N}$ is a measure of the nonlinearity of the wave.
Indeed, when this is substituted into the nonlinear equations, it can
be seen that the nonlinear terms become comparable to the linear ones
when $|{\cal N}|\exp(-k\delta)\approx1$.  Therefore the wave can be
considered to be marginally nonlinear at the surface when $|{\cal
N}|\approx1$, but, when $|{\cal N}|\approx e$, say, a substantial
fraction of the wave action can be considered to be subject to
nonlinearity.

\subsubsection{Approximate, analytical treatment of radial propagation}

To be specific, we now consider the case of objects in a circular
orbit.  (The results for non-circular orbits that provide eccentric LRs
do not differ in a substantial way from the analysis in this and the
following subsection.)  There is a Keplerian disk in which the angular
velocity is
\begin{equation}
\Omega=\left({{GM}\over{r^3}}\right)^{1/2}=\Omega_{\rm
c}\left({{r}\over{r_{\rm c}}}\right)^{-3/2},
\end{equation}
where $r_{\rm c}$ and $\Omega_{\rm c}$ are the radius and angular
velocity, respectively, of the corotation radius.  The dimensionless
intrinsic frequency of a mode with angular pattern speed $\Omega_{\rm
c}$ is
\begin{equation}
{{\hat\omega}\over{\Omega}}=m\left[\left({{r}\over{r_{\rm
c}}}\right)^{3/2}-1\right],\label{eq6}
\end{equation}
and the radii of the LRs are
\begin{equation}
r_{\rm L}=\left({{m\mp1}\over{m}}\right)^{2/3}r_{\rm c},
\end{equation}
where the upper and lower signs refer to the ILR and OLR,
respectively.  By combining equation (\ref{eq6}) with the
dimensionless dispersion relation (Figure~2), we can derive the
variation of the dimensionless wavenumber $kH$ with $r/r_{\rm c}$.
The results for various values of $m$ are shown in Figure~3.  It is
clear that, after the wave has propagated over a radial distance
comparable to $r_{\rm L}/m$ from the resonance, $kH$ has become
sufficiently large that the wave can be described accurately as a
surface gravity mode.  This makes it possible to determine the
amplitude of the wave analytically.

\placefigure{fig3}

Let $f$ denote the fraction of the torque carried by the ${\rm f}^{\rm
e}$ mode.  Then, from equation (\ref{Fa}), the flux of angular
momentum wave action associated with the wave is
\begin{equation}
\label{Tnorm}
{{\pi rm}\over{k}}\left({{\hat\omega^2-\Omega^2}\over
{\hat\omega^2}}\right)\int\rho|u|^2\,dz=\mp fT.
\end{equation}
In the limit in which the wave can be described as a surface gravity
mode, $kH$ can be approximated as $\hat\omega^2/\Omega^2$, which
itself can be considered large compared to unity, while the above
integral becomes
\begin{eqnarray}
\int\rho|u|^2\,dz&\approx&2\rho_0{\cal
N}^2\left({{\Omega^2H}\over{k}}\right)\int_0^\infty
\left({{2\delta}\over{H}}\right)^{\!\!s}e^{-2k\delta}\,d\delta\nonumber\\
&=&\Gamma(s+1){\cal N}^2(kH)^{-s}\left({{\rho_0\Omega^2H}\over{k^2}}
\right).
\end{eqnarray}
If we assume that $H\propto r$ and $\rho_0\propto r^{-p}$, the
approximate radial variation of ${\cal N}$ is given by
\begin{eqnarray}
\label{calN2}
{\cal N}^2\propto\left[\left({{r}\over{r_{\rm
c}}}\right)^{3/2}-1\right]^{2(s+3)} \left({{r}\over{r_{\rm
c}}}\right)^{p-2}.\label{eq7}
\end{eqnarray}

As the wave approaches the radial center of the disk, ${\cal
N}^2\propto r^{p-2}$.  It is quite plausible that $p>2$ (implying that
the surface density increases inwards faster than $r^{-1}$), in which
case the wave becomes {\it less} nonlinear as it travels inwards.
Otherwise, if the surface density increases inwards less rapidly,
${\cal N}^2$ increases inwards no faster than $r^{-1}$. For many cases
of interest, there is therefore nothing singular about the wave
approaching the radial center of the disk.

As the wave propagates outwards into $r\gg r_{\rm c}$, ${\cal
N}^2\propto r^{p+3s+7}$.  This means that, for any reasonable density
profile and polytropic index, the wave is very likely to become
nonlinear as it propagates outwards.

\subsection{Numerical results}

We have computed numerically the radial propagation of the ${\rm
f}^{\rm e}$ mode in a Keplerian disk using the method outlined in the
analytical treatment above, except that we do not apply the limiting
analytic forms for surface gravity waves described after equation
(\ref{Tnorm}).  The model disk has the standard parameters $s=3$,
$\gamma=5/3$, and also $H\propto r$ and $\rho_0\propto r^{-2}$.  The
results are presented as color images in Plates~1 and~2, for the cases
$m=2$ and $m=10$, respectively.  In each case, panels (a) and (b)
refer to the inward propagation from the ILR, with radii $0<r<r_{\rm
L}$ shown.  Panels (c) and (d) refer to the outward propagation from
the OLR, with radii $r_{\rm L}<r<2r_{\rm L}$ shown.  In order to show
the detail clearly, the vertical scale is greatly exaggerated.  The
two quantities plotted are the density of angular momentum wave
action, averaged over time,
\begin{equation}
{\cal
A}^{\rm(a)}=\left({{m}\over{2\hat\omega}}\right)\rho
\left(|u|^2+|w|^2\right),
\end{equation}
and the RMS velocity perturbation,
\begin{equation}
u_{\rm
RMS}=\left[{\textstyle{{1}\over{2}}}
\left(|u|^2+|v|^2+|w|^2\right)\right]^{1/2}.
\end{equation}
In panels (a) and (c), we plot the logarithm (to the base $10$) of
$|{\cal A}^{\rm(a)}|$, in units of $|T|\Omega_{\rm L}^{-1}r_{\rm
L}^{-1}H_{\rm L}^{-2}$.  In panels (b) and (d), we plot the logarithm
of $u_{\rm RMS}/(\Omega H)$, in units of $|T|^{1/2}\Sigma_{\rm
L}^{-1/2}r_{\rm L}^{-1/2}H_{\rm L}^{-3/2}\Omega_{\rm L}^{-1}$.  The
former quantity shows how the activity of the wave is distributed, in
both the vertical and the radial directions.  The latter quantity,
apart from a normalization factor, is intended as a measure of the
velocity amplitude of the wave relative to the local sound speed at
the mid-plane.  It is a more conservative measure of nonlinearity than
${\cal N}^2$, and may be more appropriate in regions close to the
resonance, where the ${\rm f}^{\rm e}$ mode behaves partly
compressibly and may produce shocks more easily.

If the disk extends continuously over the resonance, the torque can be
calculated in terms of the standard formula, equation (\ref{T}). In
many cases of interest, however, the LR truncates the disk. In such
cases, the disk edge partially overlaps with the resonance and the
torque is often physically determined by the viscosity. The density at
the resonance then adjusts so that its average value over the
resonance width provides the needed torque in accordance with equation
(\ref{T}).  The meaning of $\Sigma_{\rm L}$ in the normalization of
the velocity is the surface density expected on the basis of the
smooth power-law density profile in the main body of the disk, rather
than its actual value near the LR. In the case of a disk with an edge,
the results in Plates~1 and~2 apply where the power-law density
distribution holds in the disk.

Some care is required in interpreting Plates~1 and~2.  They are
plotted as if $H/r$ were equal to $1/2$ (for the ILR) or $1/4$ (for
the OLR), but this is done to allow a convenient visualization and has
no intrinsic significance.  The value of $H/r$ is immaterial provided
only that it is small; indeed the results become more accurate as
$H/r\to0$.  The channeling of the wave occurs on a radial scale that
is independent of $H$.  There is a region of characteristic width
$(H^2r)^{1/3}$ around the resonance where, properly, the Airy
functions should be used, but no attempt has been made to include this
on the plots because it would require a definite choice to be made for
the value of $H/r$.  In fact, the singularity of the WKB solutions at
the resonance is barely evident on the plots.

For the case $m=2$ (Plate~1), the inward propagation leads to a mild
concentration of the wave action near the surfaces of the disk.
However, the amplitude of the wave near the surface does not increase
by large amounts away from the resonance and in fact tends to a
constant value as the radial center of the disk is approached, in
accordance with equation (\ref{eq7}). The (dimensionless) amplitude
increases by about a factor of 5 from the neighborhood of the LR to
the disk center. With a larger value of $p$ (more rapid density
increase toward the center), the overall amplitude increase would be
less and the amplitude would decrease near the center.  By contrast, a
much more significant channeling effect occurs in the outward
propagation, and the amplitude near the surface increases strongly
away from the resonance, also in accordance with equation (\ref{eq7}).

For the case $m=10$ (Plate~2), the channeling effect is very
pronounced for both inward and outward propagation.  In the regions
close to the resonances, there is an approximate symmetry between the
two cases.  However, it is again seen that the amplitude tends to a
(large) constant value as $r\to0$ in the inward propagation, while the
amplitude increases strongly as $r\to\infty$ in the outward
propagation.

\section{Application to binary and protoplanetary systems}

In this section, we explore the consequences of this work for some
astrophysical situations.  In any binary star system, up to three
disks can occur.  There may be two circumstellar (CS) disks, each of
which orbits about one the stars.  In addition, there may be a
circumbinary (CB) disk that orbits around the entire system.  Powerful
tidal forces present in binary systems clear large gaps in the disks,
of order a few times the binary semi-major axis (Lin \& Papaloizou
1993; Artymowicz \& Lubow 1994, hereafter AL), which separate the
disks from each other.  As a consequence of the large size of the
gaps, the disks experience tidal forcing primarily from relatively low
azimuthal wavenumbers $m \sim 2$.

As discussed in the Introduction, there are several dissipation
mechanisms that can act to damp the waves.  This section contains some
discussion of turbulent viscous dissipation of waves.  The nature of
the disk turbulence is quite uncertain.  Additionally, the interaction
of the f mode with turbulence is complicated by the fact that the
largest turbulent eddies carry most of the viscosity, but they may not
interact efficiently to damp the waves.  Some simple-minded arguments
suggest that the spatial damping rate in the radial direction is $ k_{\rm I}
\sim \alpha k (kH)^b$, where $\alpha$ is the usual parameterization of
viscosity and $b$ is a number of small absolute value that depends
on the details of the turbulence.
%The largest eddies may be of size $H$
%which is generally larger than the wavelength of the f mode
%away from resonance. Additionally, the turnover rate of the largest
%eddies is smaller than the effective wave frequency.
Such complications add to the uncertainty in $k_{\rm I}$, but we adopt
$k_I \sim \alpha k$ as a plausible estimate.

In general, we expect wave channeling to amplify the wave
substantially over a radial length scale $r_{\rm L}/m$ for which
$kH\ga1$.  A rough relative measure of the importance of turbulent
wave damping to wave channeling is $ k_{\rm I} r_{\rm L}/m \sim \alpha
r_{\rm L}/ (m H)$.  For protostellar or protoplanetary disks, this
ratio is less than unity, while for CV disks it is greater than unity,
as is discussed below.

One important example of CB disks occurs in young binary stars, where
the remnant material involved in the star-formation process orbits
around the binary. Typically these binaries have eccentricities $e
\sim 0.3$ and the LRs are associated with the eccentric motions of the
system (AL). An LR torque is responsible for truncating the CB disk at
its inner edge.  Such CB disks have been inferred to exist from the
spectral energy distribution of some young binaries (Mathieu 1994).
Based on our results in Plate~1, we expect that significant wave
channeling would occur in a thermally stratified CB disk within a
distance $\sim r_{\rm L}/2$ from the resonance.  We expect that strong
nonlinear wave damping would occur in a CB disk, but exactly where
this would occur depends on the level of nonlinearity of the wave near
the LR. The 2D simulations of CB disks using smoothed-particle
hydrodynamics (SPH) show that the wave is somewhat nonlinear near
resonance, as is apparent from the prominence of the spiral arms near
the disk's inner edge (AL).  This effect provides some justification
for the assumption that the wave damping occurs relatively close to
the resonance (as occurs in 2D SPH simulations of CB disks, but
possibly due to artificial viscosity).

In the case of CS disks in young (eccentric) binary systems, the
eccentric Lindblad resonances can be important (GT80; AL).  As seen in
Plate~1, the wave channeling effect is relatively weak, but of
possible importance as the level of wave nonlinearity increases by
about a factor of 5 to the disk center.  For low amplitude waves in
the outer parts of the disk, the wave may well survive to the center
of the disk, provided that other damping effects can be neglected.  If
the wave reflects off the central star, then the process of wave
generation can become reversible, as the reflected wave returns to the
LR. The net effect is a zero LR torque. Other wave damping effects,
such as viscous wave damping or damping within the star, may play a
role in preventing reversibility.

In protostellar disks, the value of $\alpha$ has been estimated to be
of order $10^{-2}$ (e.g. Hartmann et al. 1998).  The extent of the
turbulent decay of a wave over radius comparable to $r_{\rm L}$ can be
estimated by taking $kH \sim m^2$, $H/r \sim 0.1$, and $m
\simeq 2$ for binary star forcing. The wave amplitude decays by
roughly a factor $ \exp{(- k_{\rm I} r_{\rm L})} \sim 0.7$.  This
plausibly provides some damping in a protostellar CS disk.  For a CB
disk, nonlinear damping by wave channeling will likely dominate, if
the wave is slightly nonlinear near resonance.

In addition, several classes of mass-transfer binaries (e.g. X-ray
binaries and CVs) possess CS disks.  These close binaries have nearly
zero eccentricity, as a consequence of the tidal interactions between
the system objects.  Under these conditions, there are no eccentric
LRs and the disk is sufficiently truncated that no (non-eccentric) LRs
can lie within the the disk.  However, the closest LR, corresponding
to $m=2$, is sufficiently powerful to generate waves off-resonance as
a result of the non-zero width of the resonance (see, e.g., Savonije,
Papaloizou, \& Lin 1994).  Again, our results suggest that wave
channeling is of possible importance, but is not very powerful.  In
practice, the vertical resonances may play a more important role in
such systems, since these resonances can lie within the disk (Lubow
1981).

CV disks surround white dwarf stars and the value of $\alpha$ has been
estimated to be of order 0.1 during outbursts, but could be much lower
during quiescence (e.g. Cannizzo 1993).  The extent of turbulent
radial decay rate over radius comparable to $r_{\rm L}$ can be
estimated by taking $kH \sim m^2$, $H/r \sim 0.02$, and $m \simeq 2$
for binary star forcing. The wave amplitude radial decay over a radius
is of order $10^8$.  These estimates are crude, but suggest that the
turbulent wave attenuation is severe (at least during outbursts), more
severe than wave channeling, and would damp the f mode.

In protoplanetary disks, the planets carve out relatively small gaps
or possibly no gap at all. As a result, several high-$m$ LRs can lie
within the disk. Under such conditions, it has been shown that the
dominant torque comes from the LR having $m \approx r/H$ (GT80; Ward
1986, 1988; Artymowicz 1993a,b). For typical protoplanetary disks, we
then expect $m \sim 10 -20$ to be most important.  Our calculations
have assumed that $m \ll r/H$, so that we cannot apply our analysis
throughout the resonant region.  In addition, these LRs have a
vertical forcing contribution that is comparable to the horizontal
forcing, and both of these vary somewhat with $z$ (Ward 1988;
Artymowicz 1993a). The present analysis assumed vertically constant
horizontal forcing and ignored some azimuthal terms in the dynamical
equations, as discussed in \S3.2.  In principle, our methods could be
extended to handle these complications.

However, even in this case, some important conclusions can be drawn
from the analysis in this paper. The region of resonant forcing in
this case can be shown to be of order $H$ in both the radial and
vertical directions (rather than $(H^2 r)^{1/3}$ for the radial and
$H$ for the vertical directions in the case $ m \ll r/H$).
Furthermore, it can be shown that the waves for which $m \approx r/H$
do approximately satisfy the equations of this paper at distances of
order a few times $H$ from the LR.  This is because the radial
wavenumber increases so rapidly with increasing distance from
resonance that it quickly dominates the azimuthal wavenumber and makes
our approximations valid.  The f mode suffers wave channeling at
radial distances from the resonance of order $r_{\rm L}/m \sim H$ in
this case.

Therefore, we are led to the simple picture that the f mode is
launched in protoplanetary disks from a region of size of order $H$.
It undergoes wave channeling on a similar scale of a few $H$.  As can
be seen from Plate~2 for $m=10$ (which we argue is approximately valid
at distances greater than about $r_{\rm L}/5$ from $r_{\rm L}$ for
$H/r \sim 0.1$), strong nonlinear damping can then occur with
increasing distance from the LR.

We expect that a full 3D analysis of the resonant region would reveal
that the f mode is still strongly excited in protoplanetary disks,
compared with other modes.  The reason is that the vertical component
of the tidal forcing is not dominant and has a single zero at
mid-plane. Furthermore, the vertical variation of the horizontal
forcing within the disk is not large, especially when the density
weighting is taken into account (Ward 1986; Artymowicz 1993a).  So the
tidal forcing should preferentially drive modes with no or few
vertical nodes. In any case, the g and r modes that are excited at an
LR undergo wave channeling that is at least as strong as that of the f
mode.

In the case of young Jupiter in the solar nebula, GT80 estimated that
the wave-induced velocities are already mildly nonlinear (having Mach
number about 0.3) at the LR.  We therefore expect that shock damping
would be important of a scale of order $H$ from resonance, where wave
channeling would occur.  The extent of turbulent wave damping on that
scale can be estimated by taking $k \sim H^{-1}$ and $\alpha \sim
0.01$.  We then obtain the level of turbulent damping $k_{\rm I} H \sim 0.1$,
and so we therefore expect that wave channeling strongly limits giant
planet driven wave propagation in thermally stratified disks.

\section{Summary and discussion}

\subsection{Summary of results}

We have analyzed the 3D response to tidal forcing of a gaseous disk
having a vertical temperature variation. The unperturbed vertical disk
structure was modeled as locally polytropic.  We have considered
effects at Lindblad resonances (LRs), subject to the thin disk
approximation $H \ll r/m$, for which the LRs provide horizontal
forcing that is independent of height.  In the standard vertically
isothermal case considered to date, the horizontally propagating 2D
mode (sound wave) carries all the resonantly generated angular
momentum, but only if the thermodynamic response of the disk is also
isothermal.

In the vertically polytropic case, the f mode has a role equivalent to
that of the 2D mode, although r and g modes are also launched at LRs
(see Figure~1). The mode launched at vertical resonances (Lubow 1981)
can be shown to be the ${\rm p}_1^{\rm e}$ mode.  We have found that
nearly all the torque exerted at an LR is carried by the f mode (see
Table~1). Near resonance, the f mode behaves in a similar manner to
the 2D mode in the isothermal case, in that it occupies the full
vertical extent of the disk and behaves somewhat compressibly.  The f
mode is almost two-dimensional near resonance in that the vertical
velocities are smaller than the horizontal by a factor of order
$(H/r)^{1/3}$.  However, away from resonance, the behaviors of the f
and 2D modes differ radically (see \S6.2).  At a distance $\sim r_{\rm
L}/m$ from the resonance, the f mode begins to concentrate its energy
near the surface of the disk, a process which we have called wave
channeling. In this regime, the f mode behaves like a surface gravity
mode (see Plates~1 and~2). In most cases, the wave amplitude increases
by many orders of magnitude with  distance from resonance, so that it
is very likely that shocks will develop which would damp the wave.  An
important exception to this last statement occurs for the $m=2$
ILR. The wave generated there undergoes relatively mild wave
channeling and consequently a relatively mild increase in nonlinearity
(by about a factor of 5). The amount
of increase depends on the density distribution in the disk and could
even result in a decrease in dimensionless amplitude near the disk
center. Dissipation by turbulent viscosity is likely 
important in some cases, such as CV disks.
On the other hand, the wave generated at the $m=2$ OLR does
exhibit strong wave channeling.  These results indicate that
circumbinary disks, as found around young binaries, and protoplanetary
disks perturbed by planets, are subject to strong effects of wave
channeling (see \S7).

\subsection{Discussion}

The results in this paper differ from the previously accepted picture
for wave propagation in thermally stratified disks (e.g. LPS). The
standard expectation was that a wave launched at a Lindblad resonance
would begin propagating horizontally, but the wavefront would be
rapidly refracted upwards as a result of the decrease in sound speed
with increasing height above the mid-plane.  After advancing a
distance comparable to the disk thickness $H$, the wavefront would be
substantially tilted upwards.  The wave would then propagate
vertically into the atmosphere of the disk, where it would shock.
This model is based on the idea that the launched wave is a pressure
wave or p mode in a high-frequency acoustic limit, so that one can
consider the wavefront to propagate at the local sound speed without
being affected by inertial or buoyancy forces.

Our results provide a different picture.  We have found that the wave
cannot be considered to propagate vertically, since it is in fact a
vertically evanescent f mode. In this view, taken in LP and KP,
the disk behaves like a waveguide in which the wave is vertically
confined. However, somewhat similar to the standard
picture, the wave energy does rise to the surface of the disk as the wave
propagates away from the resonance.  But this wave channeling process is
effective over a distance of order $r_{\rm L}/m$, where $r_{\rm L}$ is the
radius of the Lindblad resonance and $m$ is the azimuthal wavenumber.  It
does not depend on the disk thickness, provided that $H/r \ll1$. 

This can be understood roughly as follows.  The wave channeling occurs
because of the radial variation of the dimensionless intrinsic
frequency of the wave, $(\omega-m\Omega)/\Omega$.  This is equal to
$\pm\kappa/\Omega$ at the Lindblad resonance, but increases rapidly in
magnitude over a distance of order $r_{\rm L}/m$.  The wave therefore
proceeds rapidly along the f-mode branch of the dispersion relation.
In the high-frequency limit, as described in \S6.1, the mode behaves
like a surface gravity wave and is confined near the surface of the
disk.  (This would be true even in an incompressible disk in which
refraction cannot operate.)  The surface-gravity-mode dispersion
relation is approximately $(\omega -m \Omega)^2 \approx g_{\rm s} k$,
where the surface gravity $g_{\rm s}=\Omega^2 H$ and $k$ is the radial
wavenumber.  It can be seen (see \S6.1) that the mode is confined to
the disk surface in a layer of vertical thickness $\delta  \sim
k^{-1}$.  It then follows from the dispersion relation that
$(\omega-m\Omega)^2/\Omega^2 \sim H/\delta $.  As a result, the wave
becomes confined (channeled) near the disk surface (i.e., $\delta /H
\la 1/2$) over a radial distance from resonance of order $r_{\rm
L}/m$. 

Another issue concerns the boundary conditions.  In LPS, the
anticipated shocks in the disk atmosphere were represented in their
simulation by an upper boundary condition that included some amount of
dissipation.  In our analysis of a polytropic disk, we have applied a
boundary condition at the disk surface that acts to reflect waves
rather than absorb energy.  
Even in a purely isothermal disk
without a definite surface, as demonstrated by LP, all the modes are
confined vertically and do not propagate vertically to infinity.  This
is a consequence of the increase in vertical gravity with height.

Our treatment of the boundary is
 valid in the limit of high optical depth, where the
atmosphere occupies a negligible mass.  In particular, the effects
of atmosphere are not strongly felt by the wave until
$k H_{\rm atmos} > 1$, with $H_{\rm atmos}$ being
the density scale height at the base of the atmosphere. 
For a highly optically thick disk,
$H$ is much greater than $H_{\rm atmos}$, and so $k H \gg 1$.
When this condition occurs, the wave has been
strongly channeled before much wave energy enters the atmosphere.
We plan to extend our current
analysis to include the effects of a disk atmosphere in a future
paper.

The asymmetry in wave properties between the inner and outer LRs is
most pronounced for low-$m$ cases. For fixed $m$, the level of
nonlinearity for ILR waves is lower than for OLR waves (see \S6.2.2).
One reason is that the density increases as the ILR wave propagates
inward, while the density decreases as the OLR wave propagates
outwards (see eq. [\ref{calN2}]). The other cause of the asymmetry
is that the f-mode dispersion relation is followed to higher
wavenumbers in the exterior of the OLR than in the interior of the
ILR, for low $m$ (see Figure~3). Away from the LR, the f mode is
vertically confined to a surface layer of thickness $\sim k^{-1}$. Due
to wave action conservation, the confinement acts to increase the wave
amplitude, causing substantial asymmetry in the case $m=2$.  In the
high-$m$ case, the behavior of the wavenumber becomes similar near the
inner and outer LRs, and consequently the nonlinearity becomes
important at comparable distances from the respective resonances.

% Strictly speaking, the vertically polytropic model adopted here
% applies to a disk with internal heating of infinite optical depth, whereas
% the previously
% discussed 2D mode excitation applies to a disk of zero optical depth.
% Of course, for a disk of very high optical depth with internal
% heating, we expect the results of this paper to apply. To determine
% the effects of a finite optical depth, the atmosphere of the disk must
% be taken into account.  We have found a means of incorporating an
% atmosphere into our calculations and we will report on the results in
% a future paper.

A variant on the vertically polytropic model occurs for layered disks
(Gammie 1996). In this model, some regions of protostellar or
protoplanetary disks have turbulence restricted to the upper layers of
the disk. The resulting vertical temperature structure is isothermal
in the non-turbulent layer near mid-plane and the temperature declines
with height in the outer turbulent layers. A wave launched from an LR
in such a disk would probably require a longer distance from resonance
to undergo wave channeling.

Given the large increases in wave velocities associated with wave
channeling (see Plates~1 and~2), we expect that nonlinearities will
cause the waves to shock and deposit angular momentum preferentially
near the surface of the disk. We have described where nonlinearities
are expected to become important in \S6.2, based on a quasilinear
estimate. Little work has been carried out to investigate the
nonlinear behavior of f modes or surface gravity waves in a
compressible fluid. It would be useful to conduct nonlinear
simulations of the waves in such a disk.

\acknowledgments

We gratefully acknowledge support from NASA Origins of Solar Systems Grant
NAGW-4156 and NATO travel grant CRG940189. We thank Jim Pringle for many discussions on this problem.

\centerline{\epsfbox{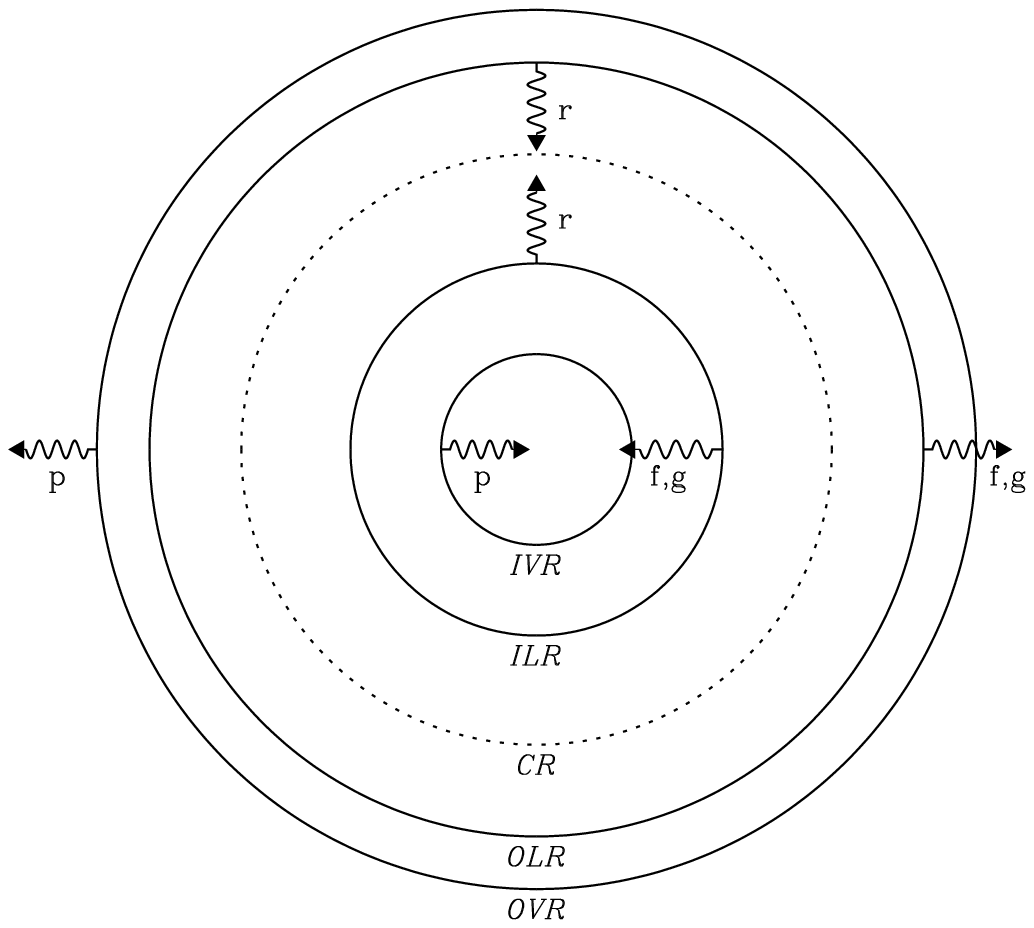}}

\figcaption[fig1.eps]{Propagation of non-axisymmetric waves in a
Keplerian disk.  The five radii marked are the corotation radius
(`CR'), the inner and outer Lindblad radii (`ILR', `OLR'), and the
inner and outer (vertical) resonances of the first p mode of even
symmetry (`IVR', `OVR').  The diagram indicates where the different
classes of modes are excited and where they propagate.  (The
parameters used here are $m=2$, $s=3$, and $\gamma=5/3$.)\label{fig1}}

\vskip1cm

\centerline{\epsfbox{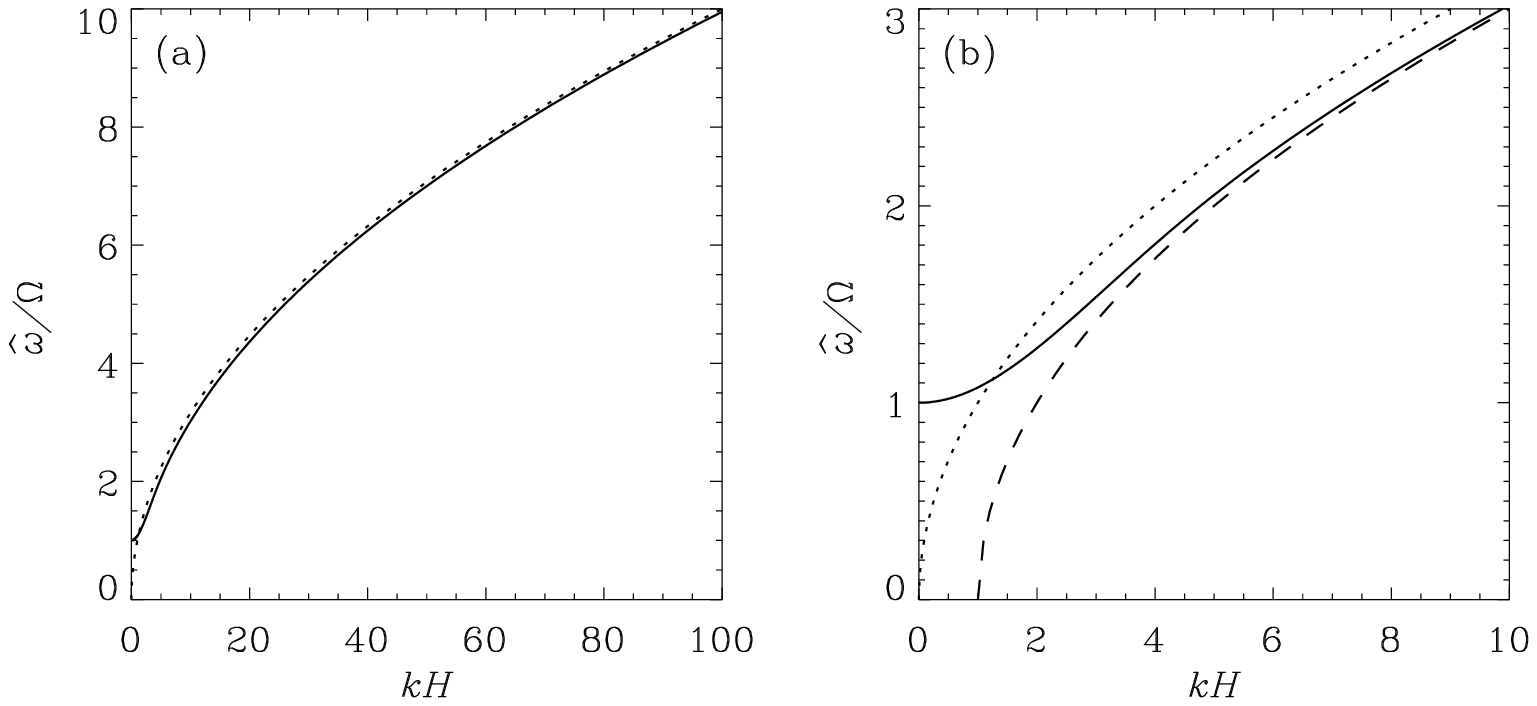}}

\figcaption[fig2.eps]{{\it Panel a\/}: dispersion relation ({\it solid
line\/}) for the ${\rm f}^{\rm e}$ mode in a Keplerian, polytropic
disk with $s=3$ and $\gamma=5/3$.  Also shown ({\it dotted line\/}) is
the asymptotic form given in equation (\ref{eq2}).  {\it Panel b\/}:
expanded view of the same dispersion relation ({\it solid line\/}).
Also shown are the approximations given in equations (\ref{eq2}) ({\it
dotted line\/}) and (\ref{eq3}) ({\it dashed line\/}).\label{fig2}}

\vskip1cm

\centerline{\epsfbox{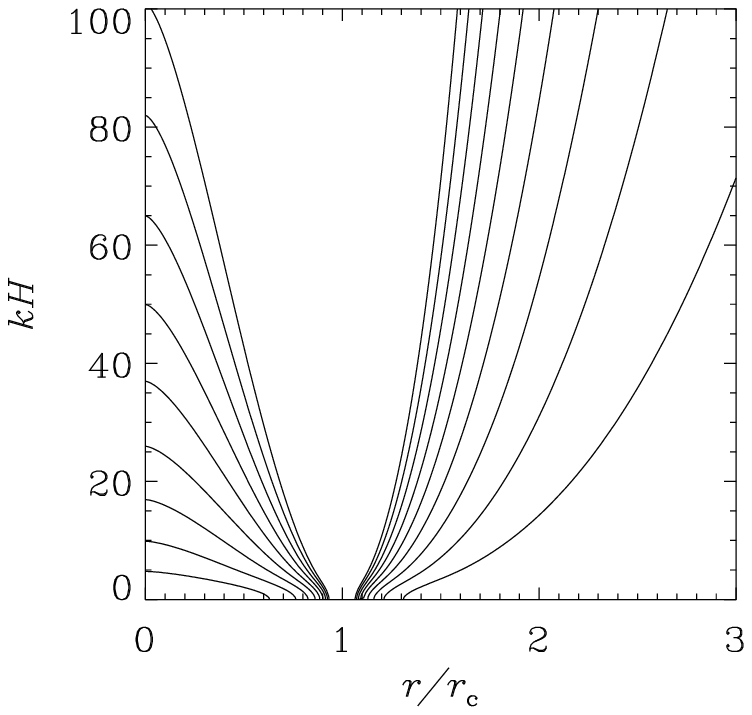}}

\figcaption[fig3.eps]{Variation of the dimensionless radial wavenumber
with radius for the ${\rm f}^{\rm e}$ mode in a Keplerian, polytropic
disk with $s=3$ and $\gamma=5/3$.  The nine curves plotted correspond
to $m=2,$ $3$, \dots, $10$, from bottom to top.\label{fig3}}

\newpage

\centerline{\epsfbox{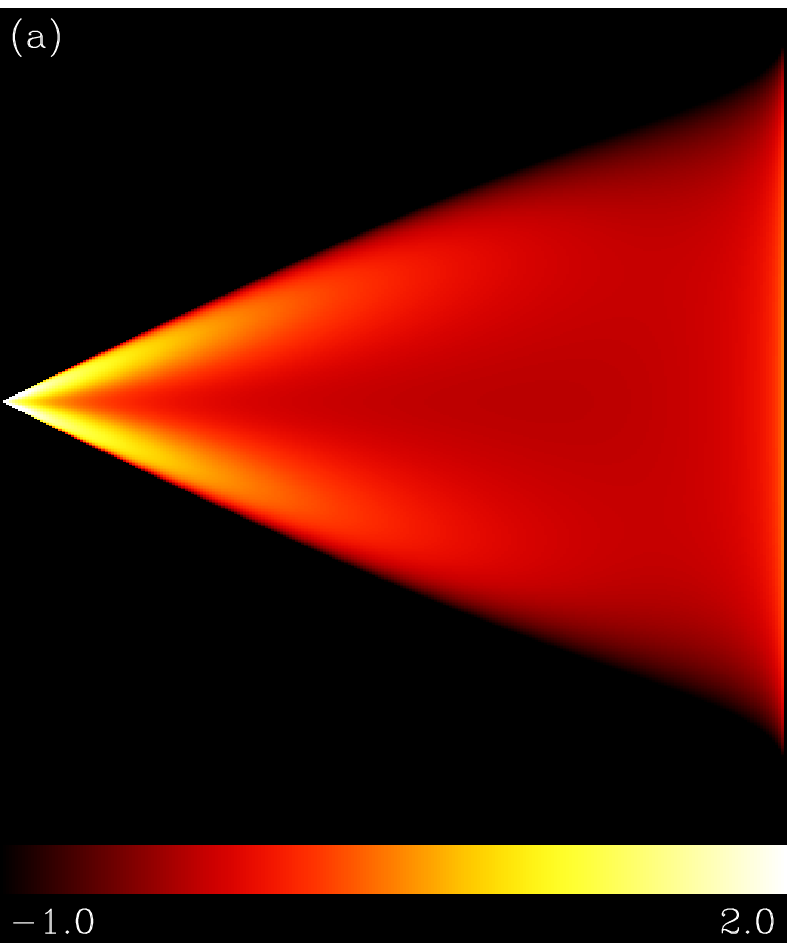}\hskip1cm\epsfbox{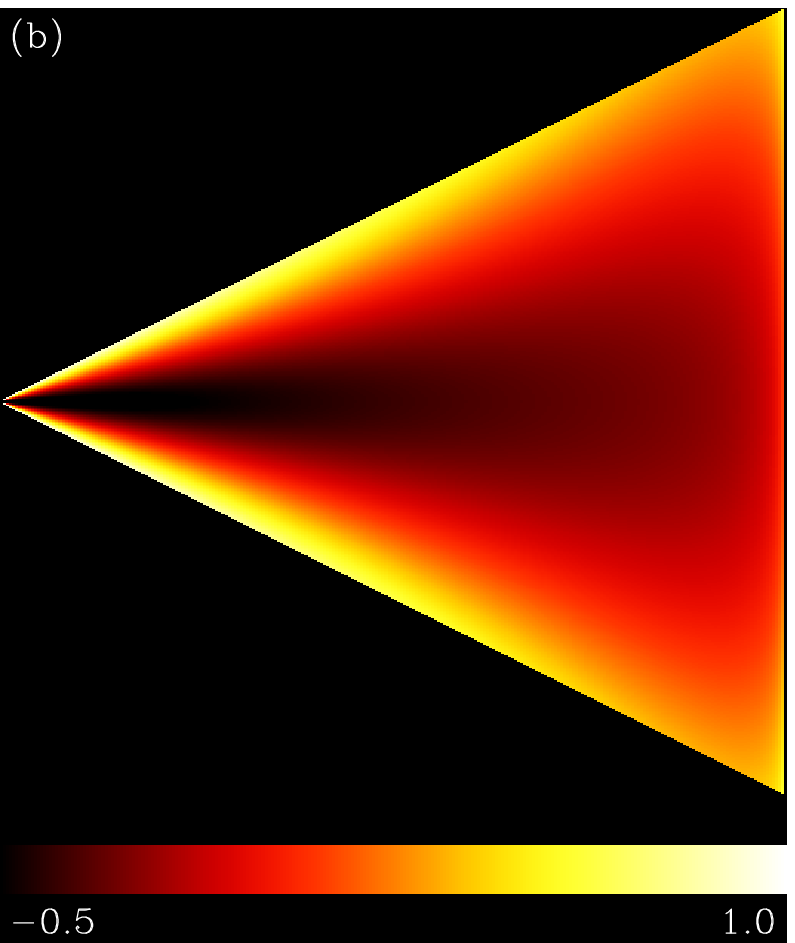}}
\vskip1cm
\centerline{\epsfbox{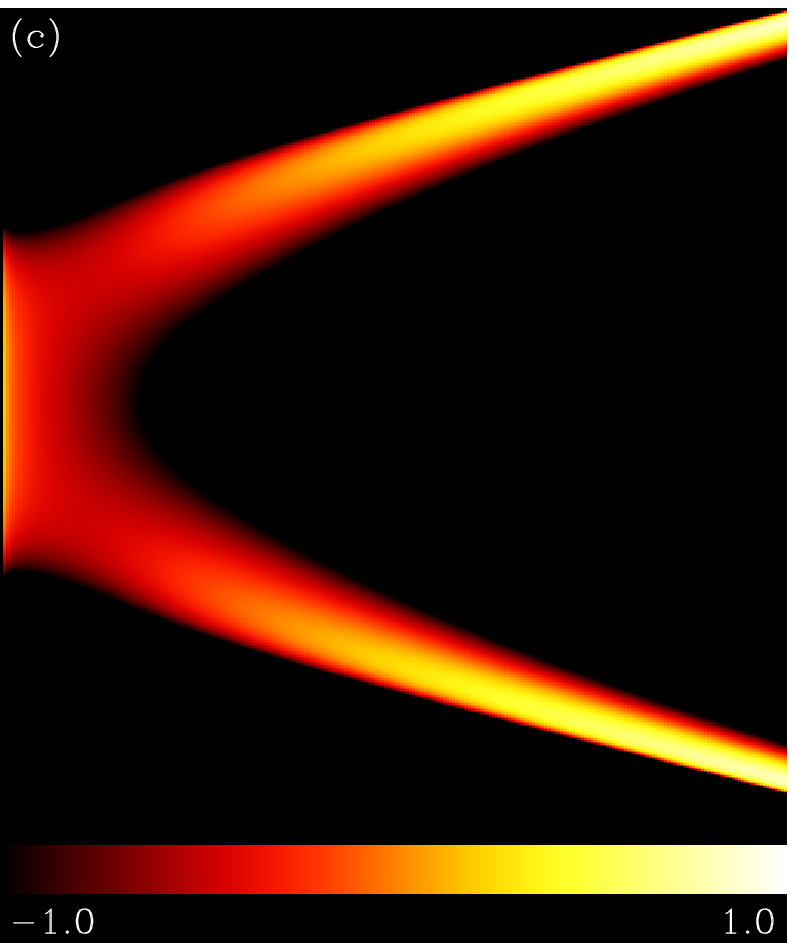}\hskip1cm\epsfbox{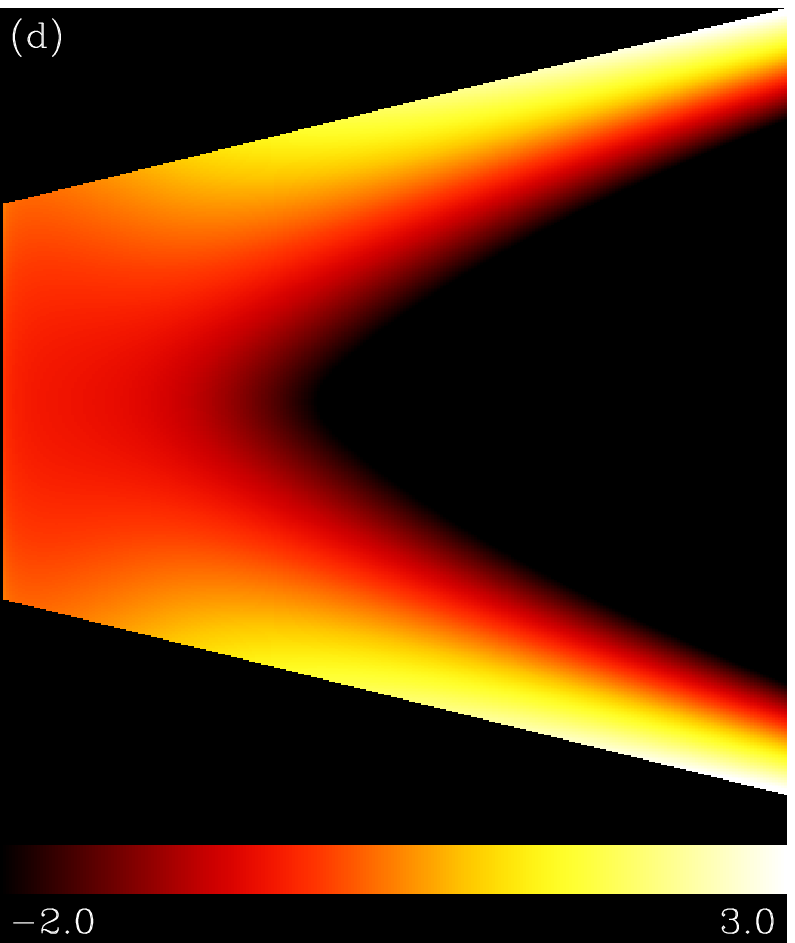}}
\vskip1cm
\leftline{\small\qquad{\sc Plate~1}.  (See Section 6.3 for discussion.)}

\newpage

\centerline{\epsfbox{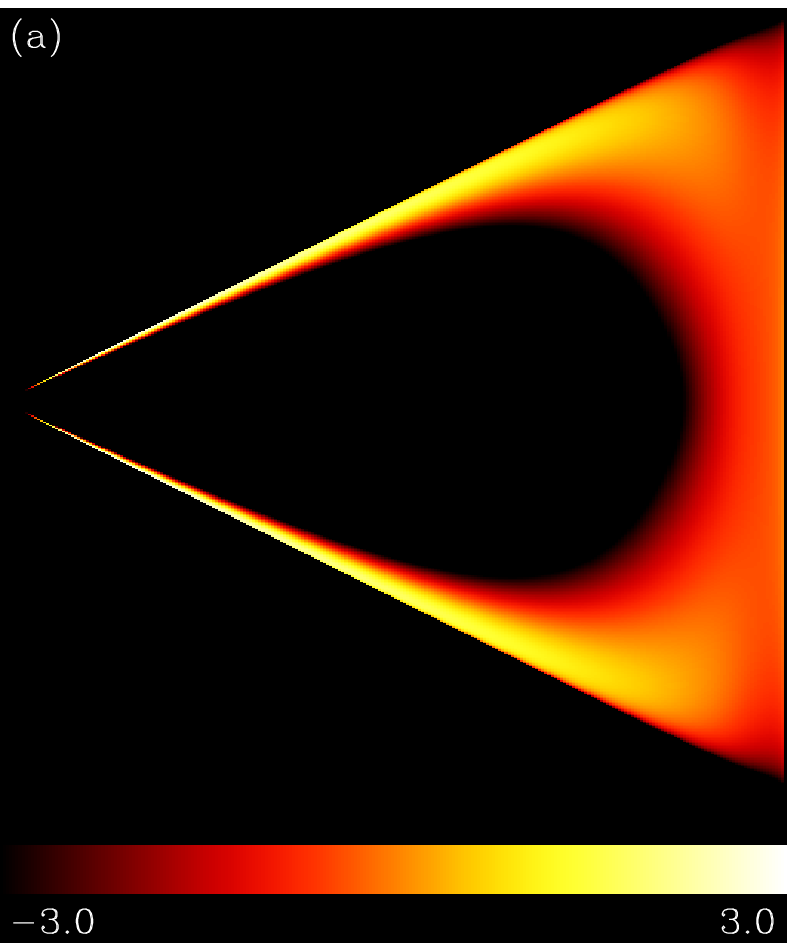}\hskip1cm\epsfbox{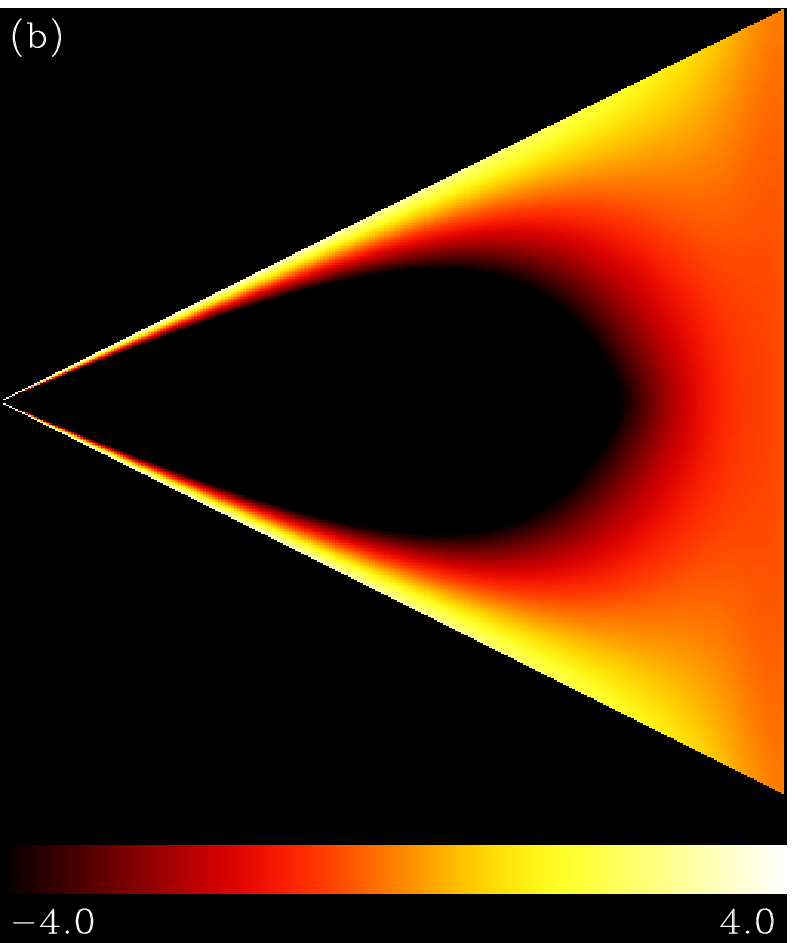}}
\vskip1cm
\centerline{\epsfbox{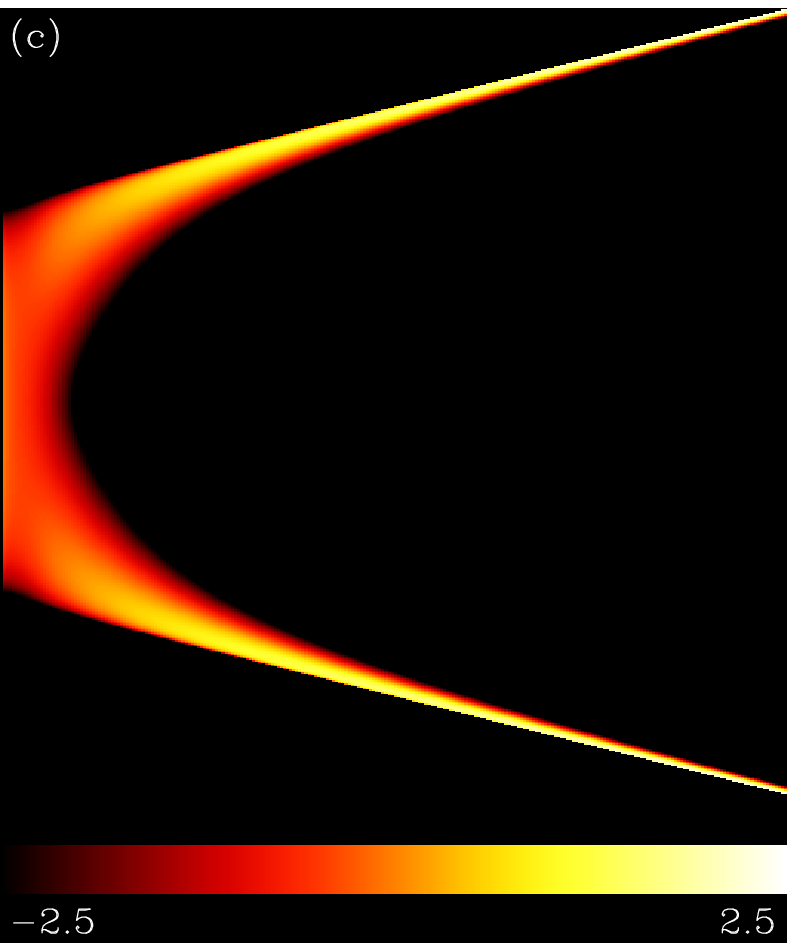}\hskip1cm\epsfbox{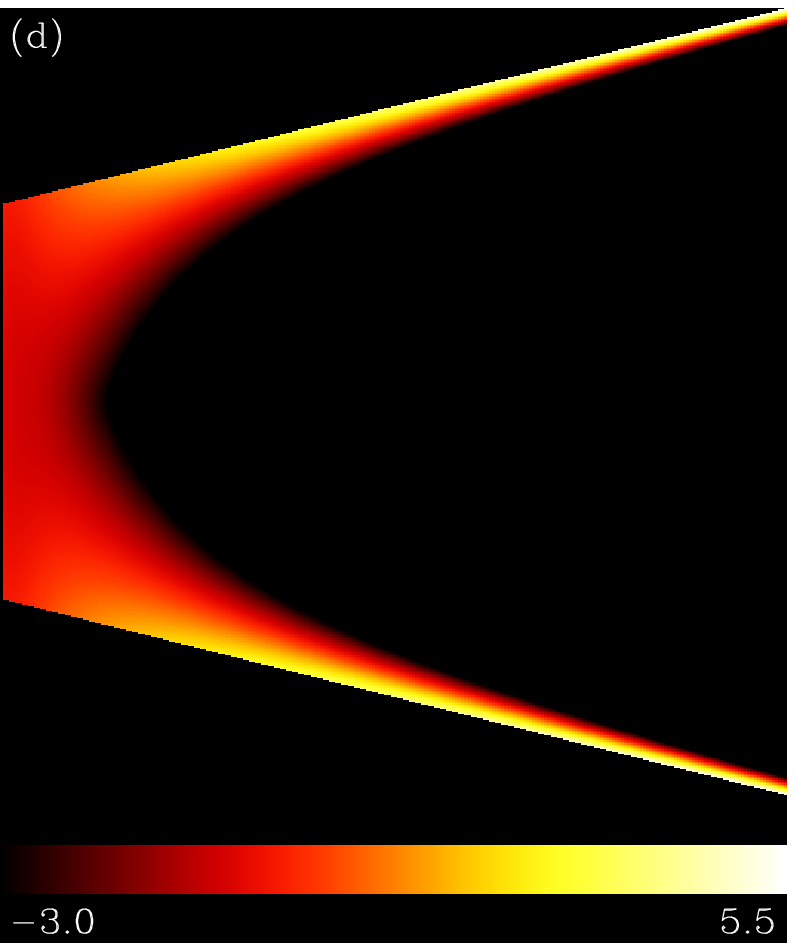}}
\vskip1cm
\leftline{\small\qquad{\sc Plate~2}.  (See Section 6.3 for discussion.)}

\end{document}